# Octave-spanning supercontinuum coherent soft X-ray for producing a single-cycle soft X-ray pulse

KAITO NISHIMIYA[1,2,5], FENG WANG[3], PENGFEI LAN[4], AND EIJI J. TAKAHASHI[1,2,*]

[1] Ultrafast Coherent Soft X-ray Photonics Research Team, RIKEN Center for Advanced Photonics, RIKEN, Wako, Japan
[2] Extreme Laser Science Laboratory, RIKEN Cluster for Pioneering Research, RIKEN, Wako, Japan
[3] Hubei Key Laboratory of Optical Information and Pattern Recognition, Wuhan Institute of Technology, Wuhan 430205, China
[4] Wuhan National Laboratory for Optoelectronics and School of Physics, Huazhong University of Science and Technology, Wuhan 430074, China
[5] e-mail: kaito.nishimiya@riken.jp
* Corresponding author: ejtak@riken.jp

**This study demonstrates the potential to generate a soft X-ray single-cycle attosecond pulse using a single-cycle mid-infrared pulse from advanced dual-chirped optical parametric amplification. A super continuum high harmonic (HH) spectrum was generated in argon (80 eV–160 eV) and neon (150 eV–270 eV). The experimental spectra reasonably agree with those calculated by the strong-field approximation model and Maxwell's equations. In addition, simulation results indicate that the dispersion of HHs in argon can be compensated using a 207-nm Zr filter to obtain 40 as pulses (1.1 cycles at 118 eV). For neon, a 278-nm Sn filter can compensate for dispersion of HH and create 23 as pulses (1.1 cycles at 206 eV). This soft X-ray single-cycle attosecond pulse is expected to be highly valuable for ultrafast science and applications in quantum information science.**

The development of intense lasers in the near infrared (NIR) to mid-infrared (MIR) region has enabled the generation and utilization of attosecond pulses in the soft X-ray (SXR) region by high harmonic generation (HHG) [1]. One atomic unit is 24 as; thus, the attosecond pulse makes it possible to observe the behavior of electrons. Although the time resolution has already reached several as [2,3], these measurements were based on the envelope of attosecond pulses (rather than the carrier) and the characteristics of the attosecond pulses have not been fully brought out. As previously demonstrated in the NIR region, it is anticipated that attosecond pulses can be utilized to observe carrier envelope phase (CEP)-dependent phenomena by reducing the number of cycles to two or less [4]. For example, focusing single-cycle attosecond pulses in the SXR region onto an atom can induce the momentum kick [5,6], which can be utilized to move electrons in an ion chain and is expected to be applicable to ultrafast control of quantum information processing [7]. Note that this phenomenon differs from tunnel ionization and multiphoton ionization [8], and it is of significant interest from a basic physical chemistry perspective. In addition, the CEP can be measured using a method that combines the interference of photoelectron energy due to one-photon ionization and two-photon ionization [4], e.g., the f-2f interferometer, which is widely used in the NIR region, or a method that combines a circularly polarized IR pulse with photoelectron momentum measurement [9].

To date, several methods have been proposed to generate a single-cycle attosecond pulse, e.g., methods based on Thomson scattering of terahertz light [10], laser plasma acceleration [11], and the free electron laser [12]. However, these methods have only been discussed theoretically and have not yet been realized. The most straightforward method to generate a single-cycle attosecond pulse is HHG. To generate a single-cycle attosecond pulse via HHG, it is essential to generate a super continuum spectrum in the cutoff region. There is a correlation between the width of this continuous region and the cycle number of the driver laser, where a shorter cycle number can generate a broader continuous HH spectrum. For example, according to calculations using SFA, when a two-cycle pulse is employed, the continuous region is approximately 15% of the entire spectrum; however, if the cycle number is single, the continuous region expands to greater than 50% [13]. Thus, to generate broader continuum HH, it is necessary to utilize the single-cycle or sub-single-cycle driver pulse.

To date, some research teams have successfully demonstrated the single-cycle attosecond pulse via super continuum HH spectrum. For example, Sansone et al. [4] generated a 1.2-cycle attosecond pulse at a photon energy of 35 eV using a sub-two-cycle pulse at a wavelength of 800 nm in combination with polarization gating (PG) [14] with Ar gas. More recently, Oguri et al. measured a broadband continuum spectrum of 70 eV centered at 120 eV with Ne harmonics using sub two cycles and double optical gating (DOG)[15, 16]. The driver wavelength is 800 nm; however, it is difficult to extend the cutoff energy up to the water window SXR, and the conversion efficiency is poor due to PG or DOG.

To generate single-cycle attosecond pulses in the SXR region, an MIR laser is required due to the wavelength-square law for HHG [17]. However, the combination with the PG or DOG is difficult due to the conversion law of the wavelength [18]. Recently, the advanced DC-OPA system was developed and it generated the pulse energy of 53 mJ, at a center wavelength of 2.44 μm while maintaining a cycle number of 1.08 [13], which is sufficient energy to compensate for the low conversion efficiency and allows the HH in the SXR region.

This paper demonstrates that a super continuum spectrum to produce single-cycle attosecond pulses at the SXR is possible using the MIR single-cycle laser generated by the advanced DC-OPA method. First, super continuum spectra were obtained in a bandwidth of one octave from 80–160 eV using Ar gas as the nonlinear medium. The continuum of the HH is greater than 50%, which allows the generation of a 1.1-cycle attosecond pulse with a Fourier transform limited (FTL) pulse of 40 as and a photon energy of 118 eV. In addition, the CEP dependence of the spectrum was compared with that of the sub-two-cycle laser, and it was found that the continuum region was clearly widened with the single-cycle laser. Furthermore, it was confirmed that the continuous region is beyond one octave by the SFA model and Maxwell's equations, and the CEP dependence of the HH spectrum agreed with the experimental values. We also generated a super continuum spectrum at higher photon energies in the SXR using Ne gas. Here, the continuous region was 150 eV–270 eV, and the FTL pulse was 22.6 as, which is below one atomic unit of time, with a central photon energy of 206 eV. The number of cycles was 1.1, and it was possible to generate single-cycle light even in the SXR region.

In this study, a single-cycle laser by the advanced DC-OPA method was employed as the laser source. The maximum laser output energy was 53 mJ with a pump energy of 700 mJ, a center wavelength of 2.44 µm, a pulse width of 8.5 fs (1.08 cycles), and a repetition rate of 10 Hz [13]. This laser adopts an intra DFG [19,20] for seed light generation; thus, a carrier-to-envelope phase (CEP) is potentially stabilized, and it can be controlled by a acousto-optic programmable dispersive filter (AOPDF), which is used for dispersion compensation. This laser system can also produce a sub-two-cycle pulse by blocking a portion of the pump laser in DC-OPA (Fig. 1). The single-cycle pulse was focused into a gas cell (1 cm) with a concave mirror (R/C = 3.0 m) to generate HH. Here, an iris was placed in front of the concave mirror, and the focusing intensity was adjusted by changing the iris size. The nonlinear medium was supplied by a pulse valve synchronized with the laser system, and a dual gas cell was employed to maintain a low pressure in the vacuum chamber [21]. The buffer cell was evacuated using a dry pump at 5000 L/min to reduce the burden on the turbo pump. The generated harmonics were observed using a flat-fields soft X-ray spectrometer with a microchannel plate. In addition, a charge-coupled device camera detected the two-dimensional fluorescence from a phosphor screen placed behind the MCP.

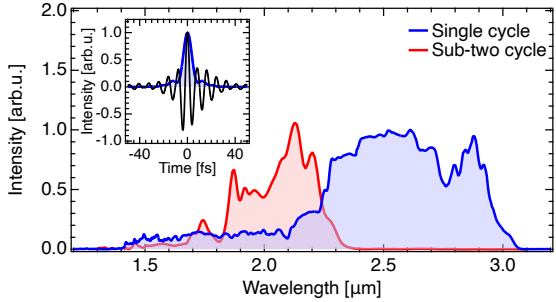

Fig. 1 Fundamental spectrum for the single (blue line) and sub-two (red line) cycle. The inset shows the temporal profile of the single-cycle laser [13].

First, HH were generated using Ar. The shot-to-shot stability in the single-shot CEP value was approximately 230 mrad (root mean squared), the CEP value of the single-cycle laser was varied by controlling the offset the AOPDF utilized for dispersion compensation, and the CEP was varied every $0.05\pi$ rad for a total of $3\pi$. The integration time of the CCD camera was 1 s, and 10 shots were integrated for each CEP value. The focused intensity was $1.3 \times 10^{14}$ W/cm$^2$, the backing pressure was 0.6 atm, and the speculated pressure in the interaction cell was approximately 20 mbar. The obtained CEP dependence of HH is shown in Fig. 2(a). As can be seen, the HH spectrum changes significantly as the CEP value changes. The signal appearing at 60 eV is a secondary diffraction signal from 120 eV, and this is an artifact of the spectrometer. A one-octave spectrum from approximately 80 eV–160 eV is a continuous region that changes depending on the CEP, and this entire region can be used as an isolated attosecond pulse. The results shown in Fig. 2(c) indicate that the half-cycle cutoff (HCO) was not observed in the HH because the phase matching condition was not satisfied. The HCO should appear on the low photon energy side (approximately half the cutoff energy). However, the phase matching condition changes for photon energy in Ar; thus, adjusting the phase matching condition near the cutoff energy makes it difficult to satisfy the phase matching condition on the low photon energy side. As a result, HCO was not observed.

For comparison, the same measurement was performed with a 1.7-cycle, 1.9-µm sub-two-cycle laser by partially blocking the pump in DC-OPA. Fig. 2(b) shows the results. Here, the focused intensity and backing pressure were maintained at the same values as that of the single-cycle laser HHG. A discrete structure was observed in the plateau region, and an HCO structure (Fig.2(c)), which typically appears when a sub-two-cycle laser is used, was also observed. In addition, the central wavelength was shifted to the shorter wavelength side; thus, the cutoff energy was shifted to the lower photon energy side.

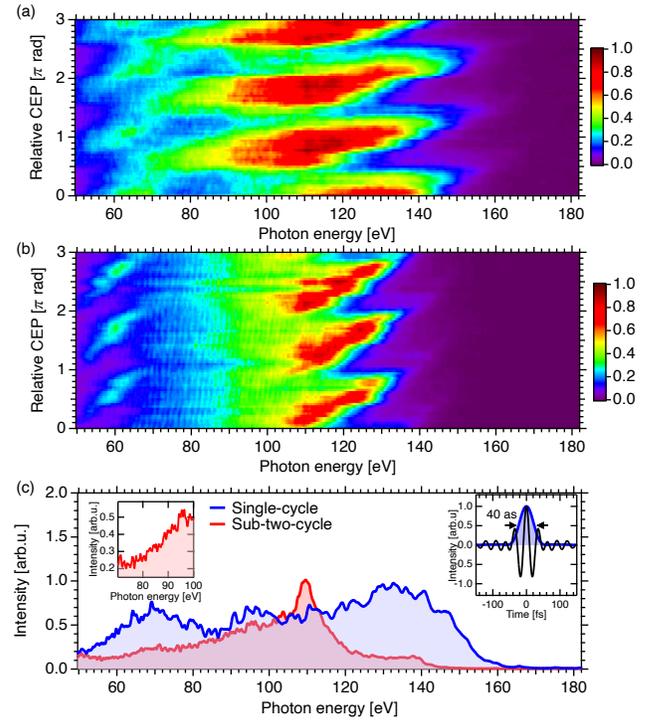

Fig. 2 CEP dependence of experimental HH spectrum by (a) a single-cycle laser and (b) a sub-two-cycle laser with Ar gas. (c) The line profile of HH spectrum at the relative CEP is $\pi$ rad.

A clear difference can be observed by comparing the CEP dependence of the HH from single-cycle and sub-two-cycle lasers. First, the position of the harmonic peak shifted as the CEP changes; however, the slope differed significantly, which is related to the change of the cutoff energy. As shown in Fig. 2, the cutoff energy changed by approximately 0.3 rad/10 eV in Fig. 2(a) but only by approximately 0.6 rad/10 eV rad in Fig. 2 (b). This difference occurred because the variance in intensity between the main peak and the second peak of the fundamental laser differs between the single-cycle and the sub-two-cycle lasers. For example, for the single-cycle laser, the value of the second peak in the electric field drops to approximately 65% of the main peak. However, for the two-cycle laser, it drops to 90%. This explains the difference in the change of the cutoff energy.

Second, in the cutoff region, the width of the continuous region that changes with CEP is wider. With the sub-two-cycle laser, the HH

spectrum contains a discrete structure at less than approximately 105 eV region (refer to the inset in Figure. 2(c)); thus, the continuous region is approximately 20%. In contrast, with the single-cycle laser, it is expanded up to approximately 50%, which is roughly in agreement with calculated value [12]. The Fourier transform of this super continuum spectrum resulted in a pulse width of 40 as with a central wavelength of 10.5 nm (photon energy: 118 eV), which corresponds to 1.14 cycles. Note that the spectrum above 80 eV was used for the calculation to eliminate the influence of artifacts due to secondary diffraction. In this calculation, atto-chirp was ignored; however, atto-chirp can be compensated.

We also compared the results with numerical calculations using SFA. Figure. 3 shows the calculation results with the single-cycle (Fig. 3(a)) and sub-two-cycle lasers (Fig. 3(b)), where the position of the photon energy peak changes when the CEP is changed, and the slope is larger with the single-cycle laser. This trend agrees with the experimental result. This corresponds to the fact that the amplitude of the electric field changes more for the single-cycle laser when the CEP is changed.

To realize single-cycle attosecond pulses, we must discuss atto-chirp. Atto-chirp is typically a positive chirp [21] that can be compensated using filters or gases [22,23]. For example, Zr transmits photon energy of 80–200 eV and gives a negative chirp of approximately −3 as/eV [14,22] with a thickness of 200 nm. The amount of chirp obtained from the SFA model and Maxwell's equations is approximately 2056 $as^2$ for a single-cycle laser, which can be compensated using a Zr filter with a thickness of 207 nm. Then, an FTL isolated attosecond pulse with a duration of 40 as can be generated.

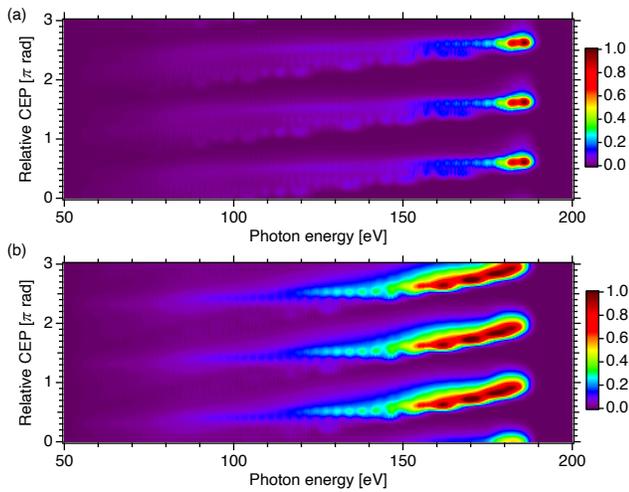

Fig. 3 CEP dependence of simulated HH spectrum by (a) single-cycle laser and (b) sub-two-cycle laser with Ar gas.

Next, the nonlinear medium was set to Ne to demonstrate that the super continuum spectrum can be generated at higher photon energy. Here, the intensity was adjusted to 2.5 x $10^{14}$ W/$cm^2$ using the iris, the backing pressure was 6.0 atm, and the speculated pressure in the interaction cell was approximately 200 mbar. The corresponding results are shown in Fig. 5. Similar to the results obtained in the Ar experiment (Fig. 2), the HH spectrum changes depending on the CEP. In Fig. 5(a), there are the three peaks in the CEP dependence of the HH. The right peak is the change in cutoff frequency, where the HH signal above 250 eV was not measured due to the low efficiency of the soft X-ray spectrometer. The center peak is the HCO, and the left peak is the secondary diffraction from the cutoff frequency. Note that the refractive index of Ne does not change significantly in the region of over 100 eV. Thus, the HH yield gradually increases as the gas pressure increases without changing the shape of the HH spectrum, which exhibits a quadratic dependence of the harmonic yield as a function of the gas backing pressure.

The continuum region changes from at least 150 eV–270 eV despite the weak HH signal at photon energies above 250 eV due to the lower diffraction efficiency. Thus, at least 40% of the HH is a continuous region. Based on the spectrum shown in Fig. 5(a), the FTL pulse in the continuous region is 22.6 as, which is less than one atomic unit of time, and the central photon energy is 206 eV. The number of cycles is 1.1 cycles, and it is possible to generate single-cycle light even in the SXR region.

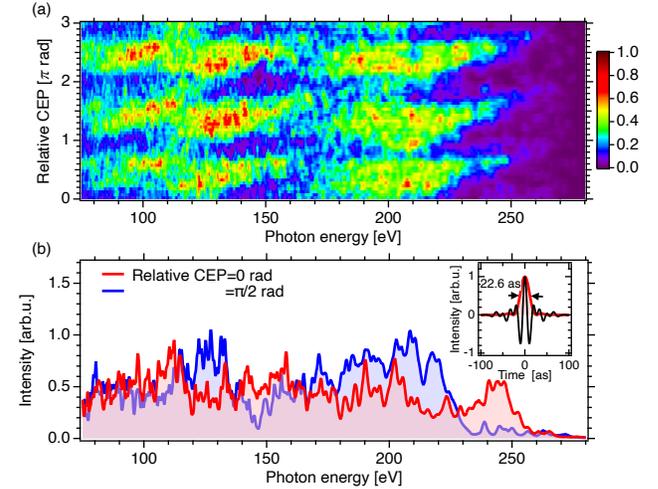

Fig 4 (a) CEP dependence of experimental HH spectrum by single-cycle laser with Ne gas. (b) Line profile of HH at the relative CEP of 0 rad and π/2 rad. The inset shows the FTL pulse calculated from the HH with a relative CEP of 0 rad.

We also compared the results with numerical calculations using the SFA model and Maxwell's equations with Ne. Figure. 5 shows the results. The HH spectrum sensitively depends on the CEP and exhibits two peaks. Here, the right peak is the change in cutoff frequency, and the left peak is the HCO. These results are consistent with the experiment results. Note that our model does not account for the secondary diffraction effect; thus, the secondary diffraction from the cutoff frequency is not observed in the numerical results.

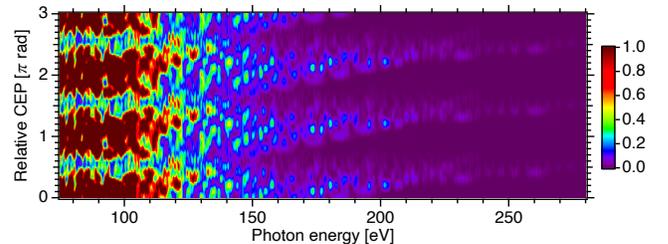

Fig. 5 CEP dependence of simulated HH spectrum by Ne.

In the 150 eV–270 eV region, the atto-chirp can be compensated by using a Sn filter. The GDD given by the Sn filter (thickness: 200 nm) is approximately 1700 $as^2$ [23]. According to our calculations, the GDD of the HH is 2368 $as^2$; thus, by using a 278-nm Sn filter, it is possible to generate single-cycle attosecond pulses in the SXR region.

In summary, this paper has demonstrated that single-cycle attosecond pulses can be generated using an MIR single-cycle laser. By changing the CEP of the fundamental laser, we measured changes over one octave, from 80 eV–160 eV for Ar and from 150 eV–270 eV for Ne, and these regions can be used as single-cycle attosecond pulses. In addition, we calculated the atto-chirp in these regions using the SFA model and Maxwell's equations, and the calculated chirp can be compensated for by optimizing the thickness of the Zr or Sn filter. These calculations confirmed that the CEP dependence of the HH spectrum is in agreement with the experiment results, especially in terms of the dependence of the HH slope on the CEP. With PG or DOG, which was used in a previous study [4,15], it is difficult to generate a single-cycle attosecond pulse in the SXR region; however, the findings of the current study have shown that it is possible to generate single-cycle attosecond pulses in the SXR region by using a single-cycle pulse as the fundamental laser. We expect that the super continuum HH generated in this study will be used for time-resolved measurements using the electric field waveform of attosecond pulses.

**Funding.** The Ministry of Education, Culture, Sports, Science and Technology of Japan (MEXT) through Grants-in-Aid under grant no. 21H01850, and the MEXT Quantum Leap Flagship Program (Q-LEAP) (grant no. JP-MXS0118068681). This project was supported by the RIKEN TRIP initiative (Leading-edge semiconductor technology).

**Acknowledgments.** We thank Dr. L. Xu for providing useful comments on handling the MIR laser.

**Disclosures.** The authors declare no conflicts of interest.

**Data availability.** Data underlying the results presented in this paper are not publicly available at this time but may be obtained from the authors upon reasonable request.

## References

1. F. Krausz and M. Ivanov, "Attosecond physics." Rev. Mod. Phys. 81, 163 (2009).
2. M. Schultze, et al., "Delay in photoemission." Science 328, 1658 (2010).
3. M. Isinger, et al., "Photoionization in the time and frequency domain." Science 358, 893 (2017).
4. G. Sansone, et al., "Isolated single-cycle attosecond pulses." Science 314, 443 (2006).
5. S. Li and R. R. Jones, "Ionization of excited atoms by intense single-cycle THz pulses." Phys. Rev. Lett. 112, 143006 (2014).
6. B. C. Yang and F. Robicheaux, "Field-ionization threshold and its induced ionization-window phenomenon for Rydberg atoms in a short single-cycle pulse." Phys. Rev. A 90, 063413 (2014).
7. H. Agueny, "Coherent electron displacement for quantum information processing using attosecond single cycle pulses." Sci. Rep. 10, 21869 (2020).
8. J. S. Briggs and D. Dimitrovski, "Ionization in attosecond pulses: creating atoms without nuclei?." New J. Phys. 10, 025013 (2008).
9. P.-L. He, et al., "Carrier-envelope-phase characterization for an isolated attosecond pulse by angular streaking." Phys. Rev. Lett. 116, 203601 (2016).
10. G. Tóth, et al., "Single-cycle attosecond pulses by Thomson backscattering of terahertz pulses." J. Opt. Soc. Am. B 35, A103 (2018).
11. Z. Tibai, et al., "Laser-plasma accelerator-based single-cycle attosecond undulator source." Phys. Rev. Lett. 114, 044801 (2015).
12. T. Tanaka, et al., "Proposal to generate an isolated monocycle X-ray pulse by counteracting the slippage effect in free-electron lasers." Appl. Phys. B 124, 113 (2018).
13. L. Xu and E. J. Takahashi, "Dual-chirped optical parametric amplification of high-energy single-cycle laser pulses." Nat. Photon. 18, 99 (2024).
14. P. B. Corkum, et al., "Subfemtosecond pulses." Opt. Lett. 19, 1870 (1994).
15. K. Oguri, et al., "Sub-50-as isolated extreme ultraviolet continua generated by 1.6-cycle near-infrared pulse combined with double optical gating scheme." Appl. Phys. Lett. 112 (2018).
16. H. Mashiko, et al., "Double optical gating of high-order harmonic generation with carrier-envelope phase stabilized lasers." Phys. Rev. Lett. 100, 103906 (2008).
17. J. L. Krause, et al., "High-order harmonic generation from atoms and ions in the high intensity regime." Phys. Rev. Lett. 68, 3535 (1992).
18. J. Tate, et al., "Scaling of wave-packet dynamics in an intense midinfrared field." Phys. Rev. Lett. 98, 013901 (2007).
19. A. Baltuška, T. Fuji and T. Kobayashi, "Controlling the carrier-envelope phase of ultrashort light pulses with optical parametric amplifiers." Phys. Rev. Lett. 88, 133901 (2002).
20. M. Zimmermann, et al., "Optical clockwork with an offset-free difference-frequency comb: accuracy of sum- and difference-frequency generation." Opt. Lett. 29, 310 (2004).
21. K. Nishimura, et al., "Apparatus for generation of nanojoule-class water-window high-order harmonics." Rev. Sci. Instrum. 92, 063001 (2021).
22. Y. Mairesse, et al., "Attosecond synchronization of high-harmonic soft X-rays." Science 302, 1540 (2003).
23. D. H. Ko, et al., "Attosecond-chirp compensation with material dispersion to produce near transform-limited attosecond pulses." J. Phys. B: At. Mol. Opt. Phys. 45, 074015 (2012).
24. Z. Chang, "Compensating chirp of attosecond X-ray pulses by a neutral hydrogen gas." OSA Continuum 2, 314 (2019).